\documentclass[twocolumn,showpacs,prb,aps,amsmath,amssymb,superscriptaddress]{revtex4}

\usepackage{graphicx}
\usepackage{dcolumn}
\usepackage{bm}
\usepackage{amssymb}

\begin{document}

\title{Nonequilibrium spin-dependent phenomena in mesoscopic superconductor-normal metal tunnel structures}

\author{Francesco Giazotto}
\email{giazotto@sns.it}
\affiliation{NEST CNR-INFM and Scuola Normale Superiore, I-56126 Pisa, Italy}
\author{Fabio Taddei}
\affiliation{NEST CNR-INFM and Scuola Normale Superiore, I-56126 Pisa, Italy}
\author{Pino D'Amico}
\affiliation{Institut f\"ur Theoretische Physik, Universit\"at Regensburg, D-93040 Regensburg, Germany}
\affiliation{NEST CNR-INFM and Scuola Normale Superiore, I-56126 Pisa, Italy}
\author{Rosario Fazio}
\affiliation{International School for Advanced Studies (SISSA), I-34014 Trieste, Italy}
\affiliation{NEST CNR-INFM and Scuola Normale Superiore, I-56126 Pisa, Italy}
\author{Fabio Beltram}
\affiliation{NEST CNR-INFM and Scuola Normale Superiore, I-56126 Pisa, Italy}


\begin{abstract}
We analyze the broad range of spin-dependent nonequilibrium transport properties of hybrid systems composed of a normal region tunnel coupled to two superconductors with exchange fields induced by the proximity to thin ferromagnetic layers and highlight its functionalities. 
By calculating the quasiparticle distribution functions in the normal region we find that they are spin-dependent and strongly sensitive to the relative angle between exchange fields in the two superconductors. 
The impact of inelastic collisions on their properties is addressed. As a result, the electric current flowing through the system is found to be strongly dependent on the relative angle between exchange fields, giving rise to a huge value of magnetoresistance. Moreover, the current presents a complete spin-polarization in a wide range of bias voltages, even in the quasiequilibrium case. In the nonequilibrium limit we parametrize the distributions with an ``effective`` temperature, which turns out to be strongly spin-dependent, though quite sensitive to inelastic collisions. By tunnel coupling the normal region to an additional superconducting electrode we show that it is possible to implement a spin-polarized current source of both spin species, depending on the bias voltages applied.    
\end{abstract}

\pacs{72.25.-b,85.75.-d,74.50.+r,05.70.Ln}

\maketitle
\section{Introduction}

Although the interest in nonequilibrium superconductivity dates back to the seventies~\cite{kopnin}, nonequilibrium transport phenomena in \emph{hybrid} superconducting structures are currently under the spotlight. One of the key experiments that renewed this interest was probably the control of the supercurrent flowing through a Josephson junction, and even the reversal of its sign, accessible by altering the quasiparticle population in the weak link (see Ref.~\onlinecite{samuelsson00} and references therein).
Out-of-equilibrium electron population can be realized in mesoscopic conductors subject to a bias voltage in which electrons cannot exchange energy either with one another or with lattice phonons, so that their energy distribution is not Fermi like~\cite{pothier}. Quasiequilibrium is reached if electrons can thermalize, while still decoupled from the phonons, so that they can reach a temperature which is different from the one relative to the phonon bath.
In ballistic Josephson junctions supercurrent control occurs by inducing a non-equilibrium population of Andreev levels either by injecting a current through an additional normal terminal connected to the weak link~\cite{vanwees,samuelsson97} or by applying an electromagnetic radiation on the weak link~\cite{shumeiko,gorelik}.
The diffusive long-junction limit was considered~\cite{volkov95,volkov97,wilhelm,yip} and experimentally realized too~\cite{morpurgo98,baselmans99}.
The control of supercurrent by cooling electrons in the weak link was proposed in Refs.~\onlinecite{giazotto03,SINIS,JT,laakso} and experimentally realized~\cite{savin04}.
It is worthwhile stressing that electron temperature can be lowered below the phonon temperature, thus realizing electron microrefrigeration \cite{SER,HT}, by exploiting the superconducting energy gap (see Refs.~\onlinecite{phystoday,RMP} and references therein).

Spin-dependent properties in out-of-equilibrium hybrid systems were investigated in a limited number of articles.
In Refs.~\onlinecite{taka,maekawa,tser,johansson} ferromagnet-superconductor-ferromagnet (FSF) double tunnel junctions were considered in order to study the spin imbalance induced in S by non-equilibrium.
In the anti-ferromagnetic alignment of the magnetizations of the F layers a strong suppression of superconductivity was found, leading to a large magnetoresistive effect.
In Josephson junctions the effect of spin injection~\cite{taka2} and presence of weak ferromagnets~\cite{bobkova} was considered, while the effect of Andreev reflection on spin accumulation in a ferromagnetic wire was reported in Ref.~\onlinecite{belzig00}.
In Ref.~\onlinecite{giazotto05} the possibility of manipulating magnetism through the interplay of superconductivity and nonequilibrium transport was investigated.
Recently we have proposed \cite{giazotto06-2} a hybrid ferromagnet-superconductor (FS) spin valve whose operation is based on the interplay between out-of-equilibrium quasiparticle dynamics and proximity-induced exchange coupling in superconductors.
Huge tunnel magnetoresistance values as high as several $10^6\%$ has been predicted, leading to a fully-tunable structure which shows high potential for application in spintronics.
In this paper we comprehensively investigate the physics and functionality of the setup analyzed in Ref.~\onlinecite{giazotto06-2}, extending our study to the presence of finite electron-electron interaction and to the quasiequilibrium limit, as well as to the presence of nonidealities in the superconductors. In this setup a spin-dependent ``effective`` temperature for the electrons in the N region emerges, thus leading to possible new spin-dependent thermoelectric effects.

The paper is organized as follows: in Sec.~\ref{setup} we describe the system under investigation and in Sec.~\ref{qp} we derive the quasiparticle distribution functions in different regimes. In particular, we consider the nonequilibrium limit in Sec.~\ref{non}, we include the effect of inelastic collisions in Sec.~\ref{sec:coll}, and we describe the quasiequilibrium regime in Sec.~\ref{sec:ql}.
In Sec.~\ref{sec:elcu} we discuss the behavior of the electric current, focusing on the magnetoresistive effects  and on the spin-filtering properties of the system in Sec.~\ref{sec:mag}.
Section~\ref{sec:te} is devoted to the characterization of the nonequilibrium distribution through an ``effective`` temperature, and to the exploitation of the system as a source of spin-polarized current through the introduction of an additional superconducting electrode.
Finally, we draw our conclusions in Sec.~\ref{sec:conc}.

\section{Setup}
\label{setup}

\begin{figure}[t!]
\includegraphics[width=\columnwidth,clip]{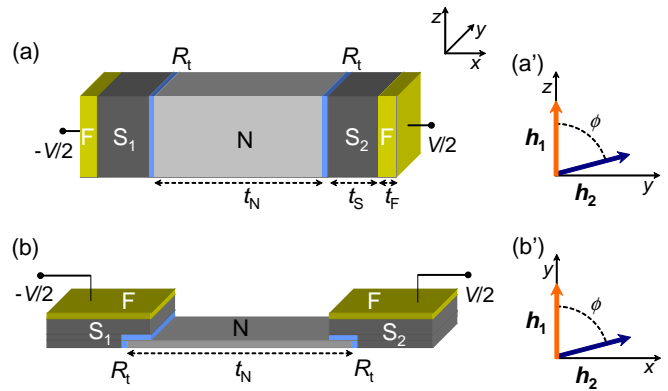}
\caption{(color online) Sketch of two possible implementations of the FS-I-N-I-SF structure analyzed in this work. (a) Spin-valve setup consisting of a sequence of stacked metallic layers. (b) A planar structure. 
Ferromagnetic layers (F) induce in each superconductor, through the proximity effect, an exchange field ($\boldsymbol{h_{1,2}}$) whose relative orientation can be controlled by an externally applied magnetic field. 
The F exchange fields are confined (a') to the $y-z$ plane for the setup shown in (a), and (b') to the $x-y$ plane for the setup shown in (b), and are misaligned by an angle $\phi$. A voltage bias $V$, applied across the structure, allows to control the energy distributions in the N region.
The structure is assumed quasi-one-dimensional.}
\label{fig1}
\end{figure}

We consider a device consisting of two identical FS bilayers (FS$_{1,2}$) symmetrically connected to a mesoscopic normal metal region (N) of length $t_{\text{N}}$ through tunnel contacts (I) of resistance $R_{\text{t}}$.
The concentration of impurities is such that quasiparticle transport is diffusive.
The resulting system, a FS-I-N-I-SF heterostructure, is shown in Fig.~\ref{fig1} in two distinct experimental implementations.
Figure~\ref{fig1}(a) shows a spin valve-like structure, which consists of a sequence of stacked metallic layers, while Fig.~\ref{fig1}(b) displays a planar system. Although the two implementations are equivalent on theoretical footing,
the planar configuration allows the measurement of local properties (e.g., the quasiparticle distribution functions as well as the local temperature) by connecting the N region to additional metallic probes. This will be addressed in Sec.~\ref{sec:te}.
For the sake of simplicity we assume a symmetric system (a resistance asymmetry would not change the overall physical picture), $t_{\text{F}}$ ($t_{\text{S}}$) labels the F (S) layer thickness and a bias voltage $V$ is applied across the structure. The exchange field in the left ferromagnet ($\boldsymbol{h_{1}}$) is aligned along the $z$ axis for the setup in Fig. \ref{fig1}(a) or along the $y$ axis for the setup in Fig. \ref{fig1}(b), while that in the right F layer ($\boldsymbol{h_{2}}$) is misaligned by an angle $\phi$ [see Fig. \ref{fig1}(a'),(b')]. For simplicity we set $|\boldsymbol{h_{1}}|=|\boldsymbol{h_{2}}|=h$.
In real structures $\boldsymbol{h_{2}}$ can be rotated by applying an in-plane magnetic field as low as some mT.
Moreover we assume that (i) the FS interface is transparent and (ii) $R_{\text{t}}$ is much larger than both the resistance of the N layer ($R_\text{N}$) and the FS contact resistance. The first condition ensures that the superconductor is strongly affected by the proximity of the F layer \cite{buzdin}, while the second ensures that all the voltage drop occurs at the tunnel barriers (so that any spatial variation of the chemical potential within the N region can be neglected), and that each FS bilayer is in local equilibrium.

The electronic properties of a FS bilayer can be analyzed within the quasiclassical Green's function formalism \cite{buzdin}. We are interested in the situation in which the influence of the F layer on the superconductor becomes nonlocal. This occurs in the limit $t_{\text{S}}<\xi_{\text{S}}=\sqrt{ \hbar D/2\pi k_{\text{B}} T_{\text{c}}}$ and $t_{\text{F}}<\xi_{\text{F}}=\sqrt{ \hbar D/h}$, where $\xi_{\text{S}}$ and $\xi_{\text{F}}$ are the superconducting coherence length and the length of condensate penetration into the ferromagnet, respectively. $D$ denotes the diffusion coefficient, $T_\text{c}$ is the superconducting critical temperature and $k_{\text{B}}$ is the Boltzmann constant.
In this situation, the ferromagnet induces in S a homogeneous \emph{effective} exchange field (analogous to the one present in magnetic superconductors \cite{buzdin}) through proximity effect and modifies the superconducting gap ($\Delta$).
The effective values of the exchange field ($h^*$) and gap ($\Delta^*$) are given by \cite{bergeret}: 
\begin{equation}
\begin{array}{l}
\Delta^*/\Delta=\nu_{\text{S}}t_{\text{S}}(\nu_{\text{S}}t_{\text{S}}+\nu_{\text{F}}t_{\text{F}})^{-1}\\
h^*/h=\nu_{\text{F}}t_{\text{F}}(\nu_{\text{S}}t_{\text{S}}+\nu_{\text{F}}t_{\text{F}})^{-1},
\end{array}
\end{equation}
where $\nu_{\text{S}}$ ($\nu_{\text{F}}$) is the normal-state density of states (DOS) in S (F).
In particular, if $\nu_{\text{F}}=\nu_{\text{S}}$ and  for $t_{\text{F}}\ll t_{\text{S}}$, it follows that
\begin{equation}
\begin{array}{l}
\Delta^*/\Delta\simeq 1\\
h^*/h\simeq t_{\text{F}}/t_{\text{S}}\ll1,
\end{array}
\end{equation}
i.e., $h^*$ turns out to be much smaller than in an isolated F layer.
As a matter of fact, $h^*$ can take values of the order of magnitude of $\Delta^*$.
These conditions can be achieved quite easily in a realistic structure.
We assume that the only effect of $h^*$ on the quasiparticles is to lead to a spin-dependent superconducting DOS, i.e., we neglect any influence of the induced magnetic moment on the orbital motion of electrons. 
Furthermore, we assume negligible spin-orbit interaction \cite{spinorbit}.
The superconductor DOS ($\mathcal{N}^{\text{S}}_{\sigma}$) thus will be BCS-like, but shifted by the effective exchange energy (equivalent to that of a Zeeman-split superconductor in a magnetic field \cite{meservey}).
By choosing the spin quantization axis along the direction of the exchange field we have
\begin{equation}
\mathcal{N}^{\text{S}}_{\sigma}(\varepsilon,h^*)=\frac{1}{2}\left|\text{Re}\left[\frac{\varepsilon+\sigma h^*+i\Gamma}{\sqrt{(\varepsilon+\sigma h^*+i\Gamma)^2-\Delta^{*2}}}\right]\right|, 
\end{equation}
where $\varepsilon$ is the energy measured from the condensate chemical potential, $\sigma =\pm 1$ refers to spin parallel (antiparallel) to the direction of $\boldsymbol{h_{1}}$, and $\Gamma$ is a smearing parameter \cite{pekola}.
The latter allows quasiparticle states within the gap due to inelastic scattering in the superconductor \cite{Dynes}, or inverse proximity effect from the nearby metallic layers. 
Typical values for $\Gamma$ lie in the range $\Gamma \sim1\times10^{-5}\Delta\ldots1\times10^{-3}\Delta$ for Al as a thin-film superconducting electrode \cite{pekola}.
In the following calculations we set $\Gamma=10^{-4}\Delta^*$, unless differently stated. 

In order to describe our system we make use of the tunneling Hamiltonian approach, and neglect proximity effects at NIS interfaces.

\begin{figure}[t!]
\includegraphics[width=\columnwidth,clip]{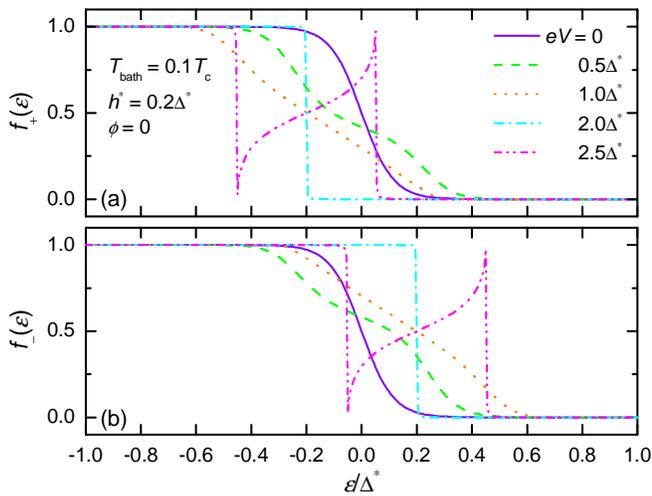}
\caption{(color online) Spin-dependent quasiparticle distribution functions $f_{\sigma}(\varepsilon)$ in the full nonequilibrium limit vs energy $\varepsilon$ for several bias voltages at $\phi=0$,  $T_{\text{bath}}=0.1T_{\text{c}}$, and $h^*=0.2\Delta^*$. (a) $f_+(\varepsilon)$; (b) $f_-(\varepsilon)$.}
\label{figfnonqePP}
\end{figure}
\section{Quasiparticle distributions}
\label{qp}
\subsection{Negligible inelastic scattering: full nonequilibrium limit}
\label{non}
At finite bias $V$ and in the limit of \emph{negligible} inelastic scattering, quasiparticles in the N layer will be out of equilibrium and thus, in general, not distributed according to the Fermi function.
The steady-state nonequilibrium distribution functions can be calculated by equating, at \emph{each} energy value, the tunneling rate of quasiparticle entering the N region from the insulating layer on the left-hand-side to the tunneling rate of those exiting through the right-hand-side barrier \cite{heslinga}.
In the general case of non-collinear exchange fields, the spin eigenstates relative to S$_2$ ($|\uparrow\rangle$ and $|\downarrow\rangle$) can be obtained by rotating the spin eigenstates relative to S$_1$ ($|+\rangle$ and $|-\rangle$) by the angle $\phi$ (representing the misalignment between $\boldsymbol{h_{1}}$ and $\boldsymbol{h_{2}}$).
As a consequence, spin up ($\sigma=+1$ with eigenstate $|+\rangle$) quasiparticles exiting the N layer through the right-hand-side barrier will now consist of two contributions.
One describes tunneling into spin up (with eigenstate $|\uparrow\rangle$) quasiparticles, proportional to $\text{cos}^2[\phi/2]$, and the other describing tunneling into spin down (with eigenstate $|\downarrow\rangle$) quasiparticles, proportional to $\text{sin}^2[\phi/2]$.
As a result, the nonequilibrium distribution function in the N layer is spin-dependent and can be written as
\begin{equation}
f_{\sigma}(\varepsilon,V,h^*,\phi)=\frac{\mathcal{N}^{\text{S}_1}_{\sigma}\mathcal{F}^{\text{S}_1}+[a(\phi)\mathcal{N}^{\text{S}_2}_{\sigma}+b(\phi)\mathcal{N}^{\text{S}_2}_{-\sigma}]\mathcal{F}^{\text{S}_2}}{\mathcal{N}^{\text{S}_1}_{\sigma}+a(\phi)\mathcal{N}^{\text{S}_2}_{\sigma}+b(\phi)\mathcal{N}^{\text{S}_2}_{-\sigma}},
\label{distribution}
\end{equation}
where $a(\phi)=\text{cos}^2[\phi/2]$, $b(\phi)=\text{sin}^2[\phi/2]$, $\mathcal{F}^{\text{S}_1(\text{S}_2)}=f_0(\varepsilon\pm eV/2)$,
$\mathcal{N}^{\text{S}_1}_{\sigma}=\mathcal{N}^{\text{S}}_{\sigma}(\varepsilon +eV/2)$, $\mathcal{N}^{\text{S}_2}_{\sigma}=\mathcal{N}^{\text{S}}_{\sigma}(\varepsilon -eV/2)$, $f_0(\varepsilon)$ is the Fermi function at bath temperature $T_{\text{bath}}$, and $e$ is the electron charge.

Figures~\ref{figfnonqePP}(a) and (b) show the nonequilibrium distributions functions [calculated from Eq. (\ref{distribution})] for spin up and spin down quasiparticles, respectively, vs energy $\varepsilon$ for the parallel configuration (i.e., $\phi=0$) at $T_{\text{bath}}=0.1T_{\text{c}}$, $h^*=0.2\Delta^*$ and different values of $V$ (we assume the superconducting gap to follow the BCS relation $\Delta^*=1.764k_{\text{B}}T_{\text{c}}$).
Figure~\ref{figfnonqePP} shows that, by increasing the bias voltage $V$, spin up and spin down distributions are shifted in opposite directions on the energy axis, similarly to what is expected in the presence of an effective spin-dependent chemical potential ($\mu_{\sigma}^{\text{eff}}$).
In particular $f_+(\varepsilon)$ is shifted toward negative energies, while $f_-(\varepsilon)$ toward positive energies. Moreover, for $eV \gtrsim \Delta^*$, the spin-dependent chemical potential saturates at $\mu_{\sigma}^{\text{eff}}=-\sigma h^*$.
As shown in Ref.~\onlinecite{giazotto05}, this effect can be used to electrostatically manipulate the magnetic properties of the N region.
The role of a finite $\Gamma$ (i.e., the presence of quasiparticle states within the gap) can be appreciated in Fig. \ref{figfnonqePP}. By increasing $eV$ from $0$ to $\Delta^*$ the distributions broaden, and reflect the ``\emph{heating}'' of the N region, as discussed in Refs.~\onlinecite{RMP,pekola}. 
This effect is absent for $\Gamma=0$. By further increasing the bias voltage the distribution functions sharpen due to ``\emph{cooling}'' provided by the superconducting energy gap \cite{RMP}.      
\begin{figure}[t!]
\includegraphics[width=\columnwidth,clip]{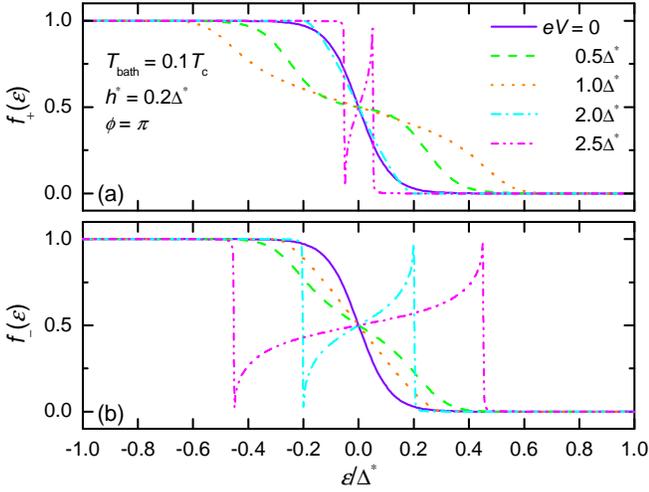}
\caption{(color online) Spin-dependent quasiparticle distribution functions $f_{\sigma}(\varepsilon)$ in the full nonequilibrium limit vs energy $\varepsilon$ for several bias voltages at $\phi=\pi$,  $T_{\text{bath}}=0.1T_{\text{c}}$, and $h^*=0.2\Delta^*$. (a) $f_+(\varepsilon)$; (b) $f_-(\varepsilon)$.}
\label{figfnonqeAP}
\end{figure}

Analogously, in Figs.~\ref{figfnonqeAP}(a) and (b) we plot the nonequilibrium distribution functions for spin up and spin down quasiparticles, respectively, for the antiparallel configuration (i.e., $\phi=\pi$). Distribution functions are shown vs energy $\varepsilon$ for different values of $V$, and were calculated for the same parameters as in Fig.~\ref{figfnonqePP}. 
In this case, up and down distributions remain centered around $\varepsilon=0$ upon biasing (equivalently, their effective chemical potential is always $\mu_{\sigma}^{\text{eff}}=0$), but at a given bias voltage the features of the distributions are more pronounced for the spin down case.
As we shall see in Sec. VI, up and down distributions are characterized by different effective electronic temperatures ($T_{\sigma}^{\text{eff}}$). 
In general, for any angle $\phi$ differing from 0 or $\pi$ the spin-dependent distribution functions $f_{\sigma}(\varepsilon)$ will be characterized by both an effective chemical potential and an effective electronic temperature.

\subsection{Intermediate inelastic scattering}
\label{sec:coll}
In the presence of scattering the approach of Sec.~\ref{non} cannot be used and one has to resort to the kinetic equation theory.
Electrons in metals experience both elastic and 
inelastic collisions.
The latter drive the system to equilibrium and can be expected to hinder the manifestation of the phenomena discussed in the previous section.
At low temperatures (typically below 1 K) electron-electron scattering \cite{alt} and scattering with magnetic impurities \cite{kaminski,anthore} are the dominant sources of 
inelastic collisions \cite{pothier,anthore,nagaev}.
Since $R_{\text{t}}$ is in general large compared to the wire resistance [$R_{\text{N}}= t_{\text{N}}/(\mathcal{N}_{\text{F}}^{\text{N}}e^2DA)$], where $\mathcal{N}_{\text{F}}^{\text{N}}$ is the N-region DOS at the Fermi energy and $A$  the wire cross-section), we can assume that $f_{\sigma}$ does not depend on the position in the wire \cite{SINIS}.

\begin{figure}[t!]
\includegraphics[width=\columnwidth,clip]{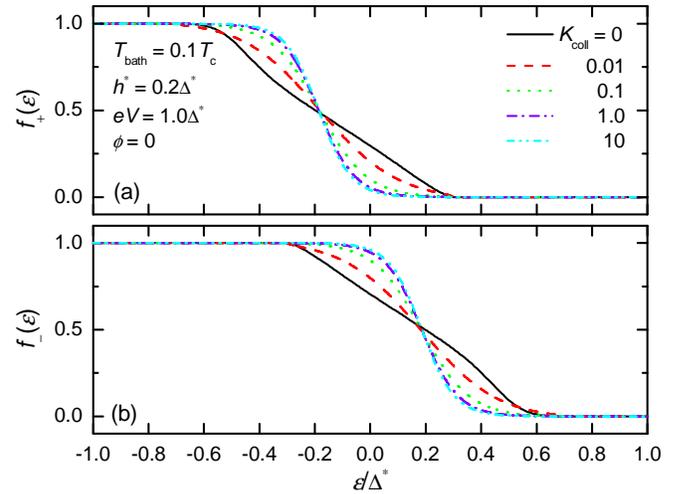}
\caption{(color online) Spin-dependent quasiparticle distribution functions $f_{\sigma}(\varepsilon)$ vs energy $\varepsilon$ calculated for several $\mathcal{K}_{\text{coll}}$ values at $\phi=0$, $eV=\Delta^{*}$, $T_{\text{bath}}=0.1T_{\text{c}}$, and $h^*=0.2\Delta^*$. (a) $f_+(\varepsilon)$; (b) $f_-(\varepsilon)$.}
\label{figfcollPP}
\end{figure}
In the following we shall analyze the role of inelastic electron-electron relaxation on the quasiparticle distribution.
The effect of electron-electron scattering due to Coulomb interaction on the spin-dependent distributions can be accounted for by solving a pair of coupled stationary kinetic equations: 
\begin{equation}\label{kinetic}
\left\{ 
\begin{split}
D\frac{\partial^2f_+(\varepsilon)}{\partial x^2}={\cal I}_{\text{coll}}^+(\varepsilon)\\
D\frac{\partial^2f_-(\varepsilon)}{\partial x^2}={\cal I}_{\text{coll}}^-(\varepsilon),
\end{split}
\right. 
\end{equation}
together with the Kuprianov-Lukichev boundary conditions at the NIS interfaces \cite{KL}.
In Eqs. (\ref{kinetic}) ${\cal I}_{\text{coll}}^{\sigma}(\varepsilon)$ is the net collision rate at energy $\varepsilon$, functional of the distributions functions $f_\sigma$, defined by 
\begin{equation}
{\cal I}_{\text{coll}}^{\sigma}(\varepsilon)={\cal I}_{\text{coll}}^{\text{in}\sigma}(\varepsilon)-{\cal I}_{\text{coll}}^{\text{out}\sigma}(\varepsilon),
\end{equation}
where
\begin{equation}\label{Icollin}
\begin{split}
{\cal I}_{\text{coll}}^{\text{in}\sigma}(\varepsilon)=
[1-f_{\sigma}(\varepsilon)]\int d\omega \frac{k(\omega)}{2}f_{\sigma}(\varepsilon-\omega)\\
\int dE \left\{ f_+(E+\omega)[1-f_+(E)]+
f_-(E+\omega)[1-f_-(E)]
\right\}
\end{split}
\end{equation}
and
\begin{equation}\label{Icollout}
\begin{split}
{\cal I}_{\text{coll}}^{\text{out}\sigma}(\varepsilon)=
f_{\sigma}(\varepsilon)\int d\omega \frac{k(\omega)}{2}[1-f_{\sigma}(\varepsilon-\omega)]\\
\int dE \left\{ f_+(E)[1-f_+(E+\omega)]+
f_-(E)[1-f_-(E+\omega)]
\right\} .
\end{split}
\end{equation}
\begin{figure}[t!]
\includegraphics[width=\columnwidth,clip]{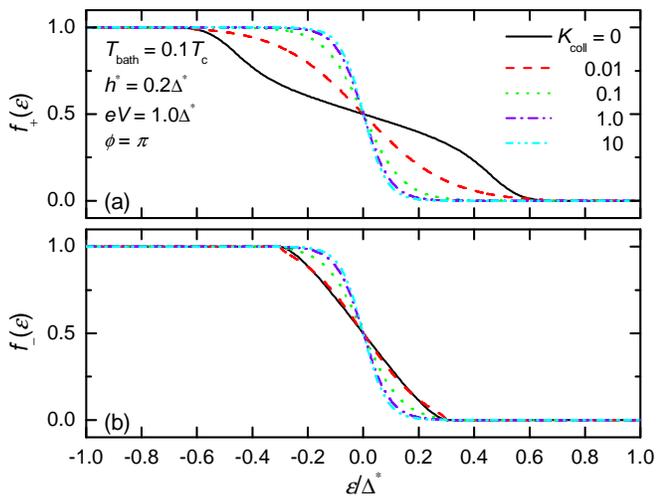}
\caption{(color online) Spin-dependent quasiparticle distribution functions $f_{\sigma}(\varepsilon)$ vs energy $\varepsilon$ calculated for several $\mathcal{K}_{\text{coll}}$ values at $\phi=\pi$, $eV=\Delta^{*}$, $T_{\text{bath}}=0.1T_{\text{c}}$, and $h^*=0.2\Delta^*$. (a) $f_+(\varepsilon)$; (b) $f_-(\varepsilon)$.}
\label{figfcollAP}
\end{figure}
In Eqs. (\ref{Icollin}) and (\ref{Icollout}), $k(\omega)=\kappa_{\text{ee}} \omega^{-3/2}$ according to the theory of screened Coulomb interaction \cite{alt2} for a quasi-one dimensional wire, where $\kappa_{\text{ee}}=(\pi\sqrt{2D}\hbar^{3/2}\mathcal{N}_{\text{F}}^{\text{N}}A)^{-1}$. \cite{kamenev,huard}
By rewriting Eqs. (\ref{kinetic}) in dimensionless units \cite{SINIS}, the strength of the electron-electron interaction can be expressed as $\mathcal{K}_{\text{coll}}=(R_{\text{t}}/R_{\text{N}})(t_{\text{N}}^2 \kappa_{\text{ee}} /D)\sqrt{\Delta}=(t_{\text{N}}/\sqrt{2})(R_{\text{t}}/R_{\text{K}})\sqrt{\Delta/\hbar D}$, where $R_{\text{K}}=h/2e^2$. We note that the strength of the electron-electron interaction turns out to be proportional to the length of the wire as well as to the tunnel barrier resistance.

We solved Eqs. (\ref{kinetic}) with $h^*=0.2\Delta^*$, $eV=\Delta^*$ and $T=0.1T_{\text{c}}$ for several $\mathcal{K}_{\text{coll}}$ values \cite{alt2,kamenev}.
The effect of electron-electron scattering on the quasiparticle distribution functions is displayed in Figs. \ref{figfcollPP} and \ref{figfcollAP} for $\phi=0$ and $\phi=\pi$, respectively. 
For the $\phi=0$ case, by increasing Coulomb interactions  the quasiparticle distributions are forced toward  thermal ones still characterized by different chemical potentials for both spin species~\cite{giazotto05}. In the antiparallel configuration (see Fig. \ref{figfcollAP}) the effect of inelastic relaxation is similar, but now the spin-dependent distribution function will coincide for sufficiently large $\mathcal{K}_{\text{coll}}$ values. It is easy to recognize that, in both cases, a thermal Fermi-like distribution is reached for $\mathcal{K}_{\text{coll}}$ of the order of 10.
Assuming parameters for a realistic Al/Al$_2$O$_3$/Ag SINIS microstructure \cite{phystoday,pekola} (with $\Delta\simeq200\,\mu$eV, $D=0.02$ m$^2$s$^{-1}$ and $R_{\text{t}}=1$ k$\Omega$) $\mathcal{K}_{\text{coll}}=10$ corresponds to a rather long N region, $t_{\text{N}}\approx 47\,\mu$m.

\subsection{Strong inelastic scattering: quasiequilibrium limit}
\label{sec:ql}
This is the regime characterized by the fact that the electron-electron interaction is so strong that quasiparticles can reach an equilibrium (Fermi-like) distribution, while the electron-phonon coupling is negligible\cite{RMP}.
Such distributions are characterized by quasiequilibrium chemical potential and temperature.
Since electron-electron interaction occurs between quasiparticles irrespective of their spin (in the absence of spin-mixing mechanisms), the quasiequilibrium temperature ($T^{\text{qe}}$) will be independent of spin, and different from the temperature of the phonon bath $T_{\text{bath}}$.
On the contrary, since electron-electron interaction redistributes the energy among electrons of a given spin species, in the absence of spin-mixing mechanisms, the quasiequilibrium chemical potential ($\mu^{\text{qe}}_{\sigma}$) will depend on spin. This is a consequence of the fact that the number of electrons of a given spin must be conserved.
Furthermore, both quasiequilibrium chemical potential and quasiequilibrium temperature will depend on $\phi$, therefore they will be different for parallel and antiparallel configurations.

In the absence of spin-flip mechanisms, the quasiequilibrium distribution functions can be calculated by imposing the conservation of \emph{particle} currents, independently for the two spin species, together with a balance equation for the \emph{heat} currents.
In particular, in the former case we require that 
\begin{equation}
I^{\text{L}}_{\sigma}(V,h^*,\phi)=I^{\text{R}}_{\sigma}(V,h^*,\phi), 
\label{qecurrent}
\end{equation}
where
\begin{equation}
I^{\text{L}}_{\sigma}(V,h^*,\phi)=\frac{1}{eR_{\text{t}}}\int d\varepsilon \mathcal{N}^{\text{S}_1}_{\sigma}(\varepsilon)[\mathcal{F}^{\text{S}_1}(\varepsilon)-f_{\sigma}(\varepsilon)]
\end{equation}
and
\begin{eqnarray}\nonumber
I^{\text{R}}_{\sigma}(V,h^*,\phi)=&\frac{1}{eR_{\text{t}}}\int d\varepsilon \left[ \mathcal{N}^{\text{S}_2}_{\sigma}(\varepsilon) a(\phi)+\mathcal{N}^{\text{S}_2}_{-\sigma}(\varepsilon) b(\phi)\right]
\\ &\times 
[f_{\sigma}(\varepsilon)-\mathcal{F}^{\text{S}_2}(\varepsilon)]
\end{eqnarray}
are the electric currents flowing through the left/right (L/R) NIS interface.
Note that, in contrast to the full nonequilibrium regime where the the tunneling rates are set to be equal at each energy, here the conservation involves the total currents, since the electron-electron interaction mixes the energy of the electrons.
In the absence of electron-phonon coupling, the only contribution to the heat flux is the heat current flowing \emph{off} the N region through each NIS interface. The latter is given by
\begin{equation}
J^{\text{L}}_{\sigma}(V,h^*,\phi)=\frac{1}{e^2R_{\text{t}}}\int d\varepsilon\; \varepsilon\; \mathcal{N}^{\text{S}_1}_{\sigma}(\varepsilon)[f_{\sigma}(\varepsilon)-\mathcal{F}^{\text{S}_1}(\varepsilon)],
\end{equation}
for the left NIS contact, and by
\begin{eqnarray}\nonumber
J^{\text{R}}_{\sigma}(V,h^*,\phi)=&\frac{1}{e^2R_{\text{t}}}\int d\varepsilon\; \varepsilon \left[\mathcal{N}^{\text{S}_2}_{\sigma}(\varepsilon)a(\phi)+\mathcal{N}^{\text{S}_2}_{-\sigma}(\varepsilon) b(\phi)\right]
\\ &\times 
[f_{\sigma}(\varepsilon)-\mathcal{F}^{\text{S}_2}(\varepsilon)]
\end{eqnarray}
for the right contact.
The balance equation for the heat flux thus simply reads 
\begin{equation}
\sum_{\sigma} \left[ J^{\text{L}}_{\sigma}(V,h^*,\phi) + J^{\text{R}}_{\sigma}(V,h^*,\phi)\right]=0.
\label{qe}
\end{equation}
By assuming that $f_{\sigma}=f_0(\varepsilon-\mu^{\text{qe}}_{\sigma},T^{\text{qe}})$ and solving  (\ref{qecurrent}) and (\ref{qe}), the temperature and chemical potentials can be easily determined.
It turns out that, while in the antiparallel alignment spin up and down distributions are equal, in the parallel one the two spin components have equal effective electronic temperature (though different from the antiparallel alignment), but opposite effective chemical potential (see Figs. \ref{figfcollPP} and \ref{figfcollAP} for large $\mathcal{K}_{\text{coll}}$ values).
Although the quasiequilibrium regime might seem an unrealistic limit, it actually describes the case of strong electron-electron interaction quite well.
Indeed, according to our calculations  (Sec. III.B), quasiequilibrium distributions are already reached  for an electron-electron collision strength ${\cal K}_{\text{coll}}\simeq10$.
In the following Sections we shall investigate the impact of quasiequilibrium on spin-dependent transport properties.

\section{Electric current}
\label{sec:elcu}

The transport properties of the FS-I-N-SF structure are determined by the spin-dependent distribution functions $f_{\sigma}$. We note that although a Josephson current can flow through the system, its theoretical description is beyond the scope of the present paper. As a matter of fact, we shall be only concerned with quasiparticle transport. Furthermore, although similar results for tunnel magnetoresistance and current polarization could be obtained in a FS-I-SF structure (i.e., without the N interlayer) and not relying on nonequilibrium, the present system possesses a crucial advantage. In fact a FS-I-SF structure implies an additional undesired Josephson current, which can be fairly large as compared to the quasiparticle current (around one order of magnitude larger than the quasiparticle current relevant for high tunnel magnetoresistance and current polarization, see for example Ref.~\onlinecite{bergeret}).  Such supercurrent could be suppressed, for instance, by the application of an additional in-plane magnetic field. This field, however, would largely exceed that required to control the orientation of $h^*$. By contrast, in the FS-I-N-I-SF system the supercurrent can be kept extremely small up to a large extent, depending mainly on $t_{\text{N}}$, on the tunnel barriers transmissivity, and on the N-interlayer material parameters. A simple estimate for the Josephson coupling in our structure reveals that the supercurrent can be from one to several orders of magnitude smaller than the quasiparticle current (see, for example, Ref.~\onlinecite{kupriyanov}).

The quasiparticle current $I$ (e.g., evaluated at the left interface) is given by 
\begin{equation}
I(V,h^*,\phi)=\sum_{\sigma}I_{\sigma}^{\text{L}}(V,h^*,\phi).
\end{equation}
\begin{figure}[t!]
\includegraphics[width=\columnwidth,clip]{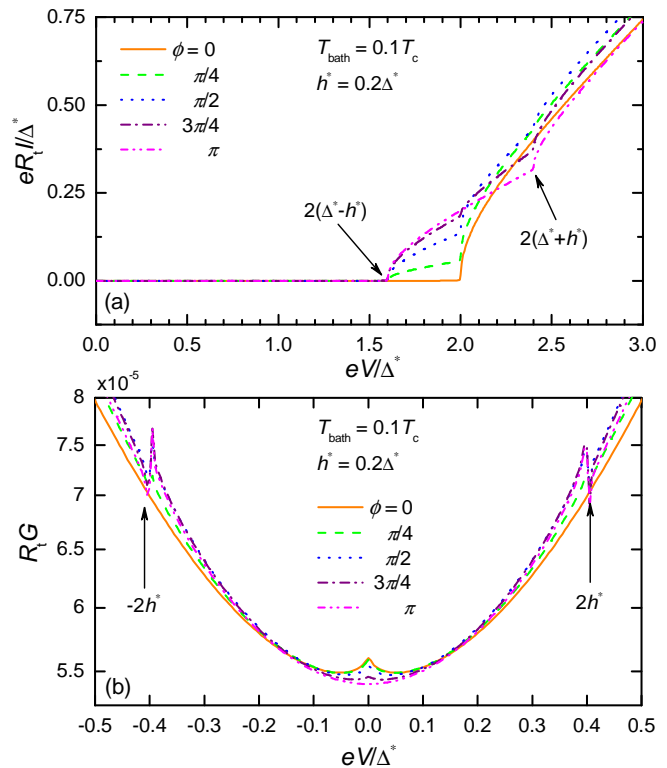}
\caption{(color online) (a) Nonequilibrium current vs bias voltage $V$ for several angles $\phi$ at $T_{\text{bath}}=0.1T_{\text{c}}$ and $h^*=0.2\Delta^*$. (b) Nonequilibrium differential conductance $G$ vs $V$ calculated for the same values as in (a). 
}
\label{tS}
\end{figure}
Figure~\ref{tS}(a) displays the electric current in full nonequilibrium vs bias voltage $V$ calculated for several angles $\phi$ at $h^*=0.2\Delta^*$ and $T=0.1T_{\text{c}}$.  
A sizable current starts to flow only when the voltage $V$ is such that the DOS is finite for both superconductors in some range of energies.
For $\phi=0$, the current rises sharply at $|eV|=2\Delta^*$, similarly to the quasiparticle current of a SIS junction (also in the presence of an in-plane magnetic field \cite{meservey}). 
In this case, in fact, the DOS of a given spin is shifted by the Zeeman energy in the same direction for both superconductors.
In contrast, for $\phi=\pi$ current sets off at $|eV|=2(\Delta^*- h^*)$.

Figure~\ref{tS}(b) shows the nonequilibrium differential conductance 
\begin{equation}
G(V,h^*,\phi)=\frac{dI(V,h^*,\phi)}{dV} 
\label{conductanceG} 
\end{equation}
calculated for the same values as in Fig.~\ref{tS}(a). Additional features are present at $|eV|=2h^*$ which are strongly temperature-dependent, and vanish in the limit $T \rightarrow 0$ (the zero-bias conductance peak for $\phi\neq\pi$ resembles that typical of a SIS junction composed of identical superconductors \cite{tinkham}).
These are a consequence of the overlapping of the superconducting DOSs where only thermally-activated quasiparticles exist at finite temperature.

\begin{figure}[t!]
\includegraphics[width=\columnwidth,clip]{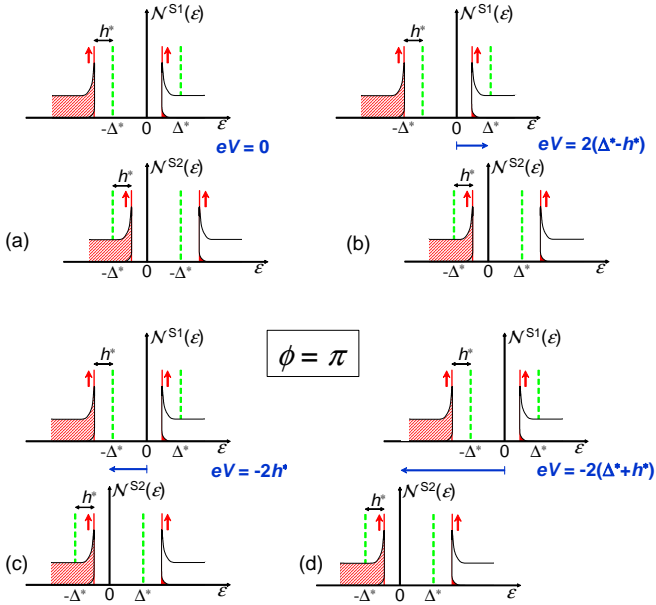}
\caption{(color online) Idealized finite-temperature exchange field-split density of states $\mathcal{N}^{\text{S}_{1,2}}$ of S$_{1,2}$ for parallel spin species, at $\phi=\pi$ and different bias voltage $V$. In particular, (b), (c) and (d) show how features in the tunneling current originate at $eV=2(\Delta^*-h^*)$, $eV=-2h^*$, and $eV=-2(\Delta^*+h^*)$, respectively. Antiparallel spin species gives rise to features at opposite voltages. Green-dashed lines represents the superconducting DOS in the absence of the exchange field.
}
\label{schemeDOS}
\end{figure}

All this simply reflects how the spin-dependent DOS in each superconductor contributes to the total quasiparticle current at different $V$. This can be easily visualized by inspecting Fig. \ref{schemeDOS}, which shows  idealized finite-temperature exchange field-split superconducting DOS for parallel spin species, at different bias voltage $V$ and for the case $\phi=\pi$. 
In this case the DOS of S$_{1}$ is shifted in the opposite direction with respect to that of S$_2$ [see Fig. \ref{schemeDOS}(a) for $V=0$]. Then, by biasing the structure, the required voltage for a current to flow is smaller with respect to the $\phi=0$ case, i.e., $eV=2(\Delta^*-h^*)$ [see Fig. \ref{schemeDOS}(b)]. In the same way, for negative voltages, the current sets off at $eV=-2(\Delta^*+h^*)$, as shown in Fig. \ref{schemeDOS}(d).
It is also clear that antiparallel spin species will give rise to features at the opposite bias voltage, therefore explaining the origin of additional feature appearing at $eV=2(\Delta^*+ h^*)$.
For intermediate values of $\phi$, features are present at $|eV|=2(\Delta^*\pm h^*)$ and at $|eV|=2\Delta^*$, since contributions from both $\phi=0$ and $\phi=\pi$ configurations are present.
Of particular relevance is the voltage interval $2(\Delta^*-h^*)\leq |eV|\leq 2\Delta^*$. By increasing $\phi$ from $0$ to $\pi$, the current is enhanced from a vanishingly small value up to a finite value leading to a \emph{spin-valve} effect.

It is noteworthy to mention that the nonequilibrium condition is essential for the observation of the spin-valve effect. At equilibrium the distribution functions in the N layer would be thermal and spin-independent.
\begin{figure}[t!]
\includegraphics[width=\columnwidth,clip]{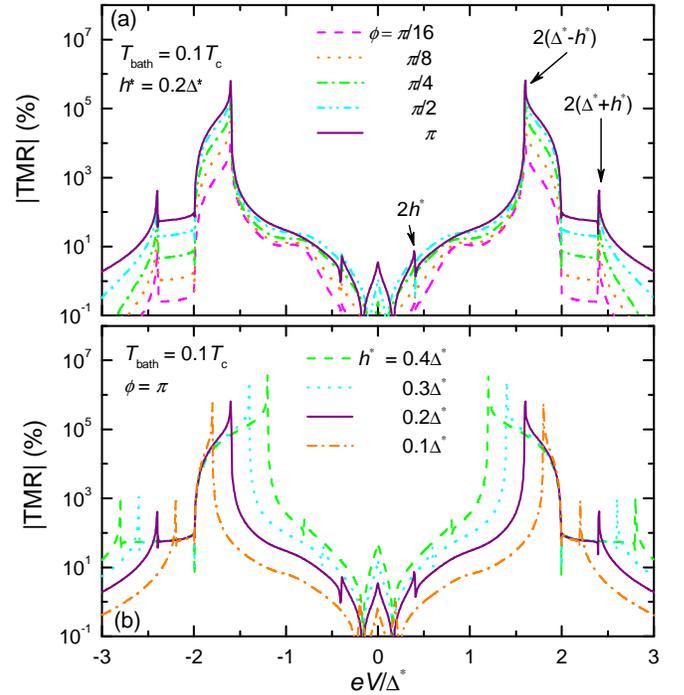}
\caption{(color online) (a) Nonequilibrium tunnel magnetoresistance ratio $|\text{TMR}|$ vs $V$ calculated for several angles $\phi$ at $T_{\text{bath}}=0.1T_{\text{c}}$ and $h^*=0.2\Delta^*$. (b) $|\text{TMR}|$ vs $V$  for different $h^*$ values at $T=0.1T_{\text{c}}$ and $\phi=\pi$.}
\label{tF}
\end{figure}

\section{Magnetoresistance}
\label{sec:mag}

The spin-valve properties of the FS-I-N-I-SF setup can be evaluated quantitatively by analyzing the tunnel magnetoresistance ratio ($\text{TMR}$), defined as
\begin{equation}
\text{TMR}(V,h^*,\phi)=\frac{G(V,h^*,\phi)-G(V,h^*,0)}{G(V,h^*,0)}.
\end{equation}
Figure~\ref{tF}(a) displays the absolute value of the nonequilibrium TMR vs bias voltage $V$ calculated for several angles $\phi$ at $T=0.1T_{\text{c}}$ and $h^*=0.2\Delta^*$.
For $2(\Delta^*-h^*)\leq |eV|\leq 2\Delta^*$ the TMR increases monotonically by increasing $\phi$ and is maximized at $\phi=\pi$ where it reaches huge values exceeding $10^6\%$. We note that in the limit $T=0$ and $\Gamma=0$, $|\text{TMR}|$ diverges, realizing an ideal full spin-valve effect.  
The nonequilibrium TMR behavior for several  exchange field values is shown in Fig.~\ref{tF}(b), at $T=0.1T_{\text{c}}$ and $\phi=\pi$. By decreasing $h^*$, the maximum TMR value reduces, and so does the voltage interval of larger magnetoresistance.  Larger $h^*$ values are thus preferable in order to extend the voltage window for optimized operation and to maximize the TMR. 

\begin{figure}[t!]
\includegraphics[width=\columnwidth,clip]{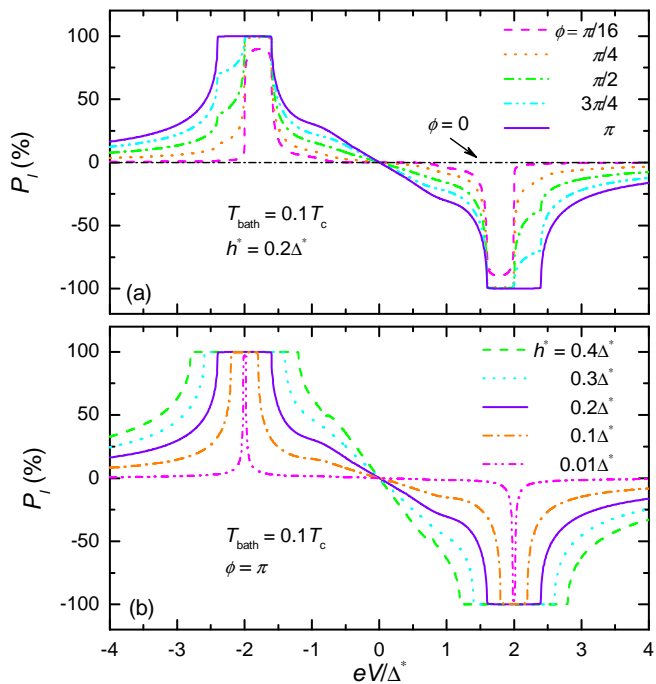}
\caption{ (color online) (a) Nonequilibrium current polarization $P_I$ vs $V$ calculated for several angles $\phi$ at $T_{\text{bath}}=0.1T_{\text{c}}$ and $h^*=0.2\Delta^*$. (b) $P_I$ vs $V$ for different $h^*$ values at $T_{\text{bath}}=0.1T_{\text{c}}$ and $\phi=\pi$.}
\label{exch}
\end{figure}

The spin-filtering properties of this system can be quantified by inspecting the current polarization ($P_I$), defined as
\begin{equation}
P_I(V,h^*,\phi)=\frac{I_{+}^{\text{L}}(V,h^*,\phi)-I_{-}^{\text{L}}(V,h^*,\phi)}{I_{+}^{\text{L}}(V,h^*,\phi)+I_{-}^{\text{L}}(V,h^*,\phi)}.
\end{equation}
The calculated nonequilibrium $P_I$ vs $V$ is displayed in Fig.~\ref{exch}(a) for several $\phi$ values, at $T=0.1T_{\text{c}}$ and $h^*=0.2\Delta^*$. Upon increasing $\phi$, two intervals of $100\%$ spin-polarized current develop for $2(\Delta^*-h^*)\leq |eV|\leq 2\Delta^*$, extending to wider regions [$2(\Delta^*-h^*)\leq |eV|\leq 2(\Delta^*+h^*)$] as $\phi$ approaches $\pi$. For $\phi=0$, $P_I$ vanishes like in SIS junctions with an in-plane magnetic field \cite{meservey}.
Depending on bias, fully spin-polarized currents of both parallel and antiparallel spin species can be obtained. The structure can thus be also operated as a \emph{controllable} spin-filter by changing the orientation of $\boldsymbol{h_{2}}$ as well as by varying $V$.
Figure~\ref{exch}(b) shows $P_I$ vs $V$ for several $h^*$ at $T=0.1T_{\text{c}}$ and $\phi=\pi$. The net effect of increasing $h^*$  is to widen the regions of $100\%$ spin-polarized current.

\begin{figure}[t!]
\includegraphics[width=\columnwidth,clip]{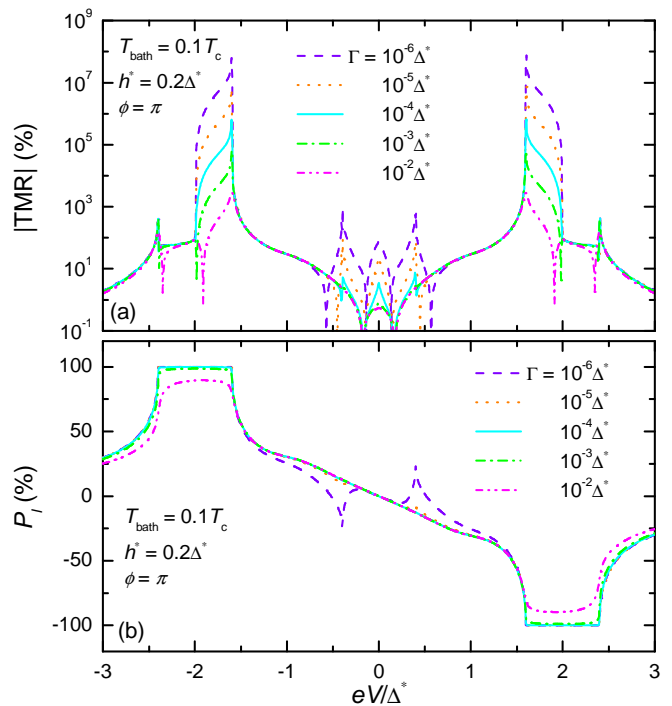}
\caption{ (color online) (a) Nonequilibrium tunnel magnetoresistance ratio $|\text{TMR}|$ vs $V$ calculated for several $\Gamma$ values at $T_{\text{bath}}=0.1T_{\text{c}}$, $h^*=0.2\Delta^*$, and $\phi=\pi$. (b) Nonequilibrium $P_I$ vs $V$ calculated for the same values as in (a).}
\label{gamma}
\end{figure}

It is important to discuss the effect of the smearing parameter $\Gamma$ (which controls the presence of quasiparticle states within the superconducting gap) on the magnetoresistance and current polarization.
As shown in Fig.~\ref{gamma}(a), by increasing $\Gamma$, the TMR value decreases mostly in the region $2(\Delta^*-h^*)\leq |eV|\leq 2\Delta^*$, while for other values of $V$ almost no changes are found apart from some smoothing of sharp features.
In particular, the normal character of transport is strengthen by increasing $\Gamma$ which causes a suppression of the large TMR value. The latter indeed is a consequence of the presence of the superconducting gap.
On the contrary, the impact of $\Gamma$ on $P_I$, plotted in Fig.~\ref{gamma}(b) as a function of the voltage $V$, is much weaker: the polarization in the range $2(\Delta^*-h^*)\leq |eV|\leq 2(\Delta^*+h^*)$ is almost insensitive to $\Gamma$, being slightly reduced only for $\Gamma$ values as large as $10^{-2}\Delta^*$.

\begin{figure}[t!]
\includegraphics[width=\columnwidth,clip]{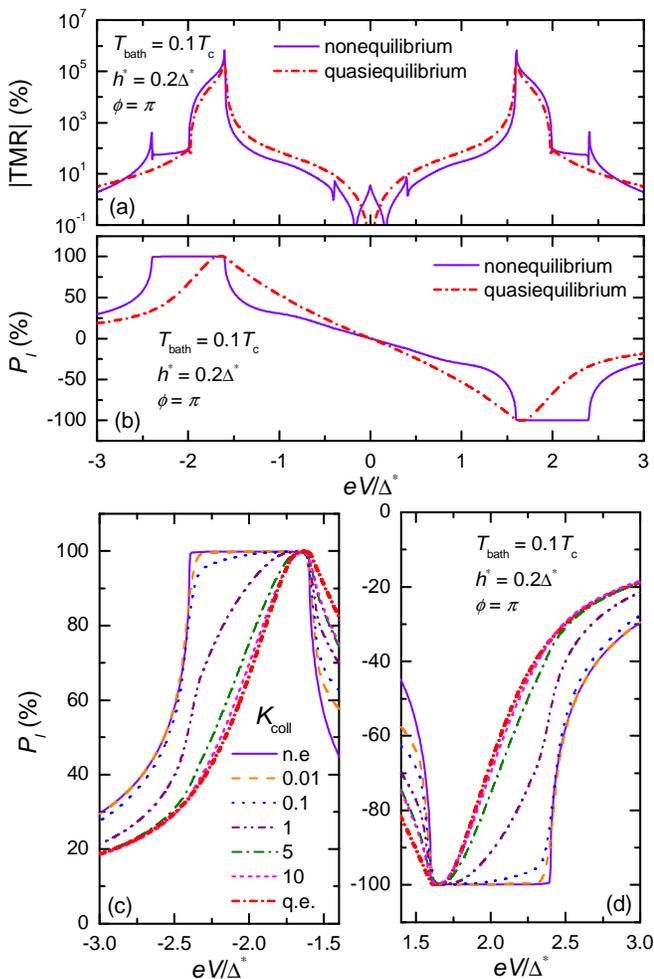}
\caption{ (color online) (a) Tunnel magnetoresistance ratio $|\text{TMR}|$ vs $V$ in nonequilibrium (solid line) and quasiequilibrium (dash-dotted line) at $T_{\text{bath}}=0.1T_{\text{c}}$, $h^*=0.2\Delta^*$, and $\phi=\pi$. (b) Current polarization $P_I$ vs $V$ in nonequilibrium (solid line) and quasiequilibrium (dash-dotted line) calculated for the same values as in (a). (c-d) $P_I$ vs $V$ in the intervals of large polarization calculated for several $\mathcal{K}_{\text{coll}}$ values and the same parameters as in (a).
}
\label{quasiequilibrium}
\end{figure}

TMR values are expected to be marginally affected by the presence of electron-electron relaxation in the N layer.
Indeed, as discussed in Secs.~\ref{sec:coll} and \ref{sec:ql}, Coulomb interaction allows quasiparticles to exchange energy (through inelastic collisions) without coupling the two spin species.
As shown in Figs.~\ref{figfcollPP} and \ref{figfcollAP}, inelastic scattering leaves the two distributions $f_+(\varepsilon)$ and $f_-(\varepsilon)$ strongly spin-dependent in the parallel configuration, while making them to coincide in the antiparallel configuration, so that both magnetoresistance and current polarization are expected to be only slightly affected.
Indeed, TMR is only marginally affected even in the quasiequilibrium regime, as shown in Fig.~\ref{quasiequilibrium}(a), where we compare TMR at $\phi=\pi$, as a function of $V$, for the full nonequilibrium and the quasiequilibrium regimes.
The effect of energy redistribution characteristic of quasiequilibrium consists merely in a smoothing of some of the sharp features present in the nonequilibrium limit.
In Fig.~\ref{quasiequilibrium}(b), we compare the plots of $P_I$ at $\phi=\pi$ as functions of $V$ for both regimes.
In particular, quasiequilibrium displays a reduction of polarization for $|V|>2(\Delta^*-h^*)$, and an increase of polarization for $|V|<2(\Delta^*-h^*)$. Nevertheless, $P_I$ values as large as $100\%$ can be obtained in the quasiequilibrium limit as well.

The full evolution of $P_I$  as a function of $\mathcal{K}_{\text{coll}}$ in the relevant intervals of large polarization  is shown in Fig.~\ref{quasiequilibrium} (c-d). It is easy to notice the gradual smearing of $P_I$ by increasing the electron-electron interaction strength, and that the quasiequilibrium limit is already reached  for $\mathcal{K}_{\text{coll}}\sim 10$ (see also the discussion at the end of Sec.~\ref{sec:coll}). However, in FS-I-N-I-SF realistic structures it should be possible to keep $\mathcal{K}_{\text{coll}}\sim 0.1$ or smaller so that current polarization would be somewhat similar to that obtained in the full nonequilibrium limit. 

By contrast, TMR must decrease if spin-flip processes mix the spin-dependent distributions. In metals and at low temperature (typically below $\sim1$ K), such processes are normally caused by the presence of magnetic impurities in the N layer. 
Spin-flip scattering can be suppressed by limiting the magnetic-impurity content in the N layer, and by choosing $t_{\text{N}}\ll \lambda_{\text{sf}}$ (the spin-flip relaxation length $\lambda_{\text{sf}}$ is of the order of some $\mu$m in metals such as Cu or Au \cite{jedema,johnson}). These constraints can be met fairly easily experimentally in multilayered or planar structures like the ones presented here.

\section{Spin-dependent effective temperature}
\label{sec:te}
Even in the nonequilibrium case, it is interesting to characterize the distribution function through an ``effective'' temperature and an ``effective'' chemical potential.
Such effective quantities can be meaningfully defined by fictitiously connecting the N region to a large normal-metal reservoir through an insulating barrier that is sufficiently transparent to allow the flow of quasiparticles, but opaque enough not to alter the nonequilibrium condition of the electrons.
One can identify the effective chemical potential ($\mu^{\text{eff}}_{\sigma}$) of the out-of-equilibrium electron gas in the N region with the chemical potential the reservoir must possess in order for the particle current to be zero.
The effective temperature ($T^{\text{eff}}_{\sigma}$), on the other hand, is taken to be equal to the one the reservoir must have in order for the heat current to be zero \cite{tesiheikkila}.
These conditions can be expressed respectively by the following two equations:
\begin{eqnarray}
\int d\varepsilon [f_{\sigma}(\varepsilon)-f_0(\varepsilon-\mu^{\text{eff}}_{\sigma},T^{\text{eff}}_{\sigma})]=0\\
\int d\varepsilon \; \varepsilon [f_{\sigma}(\varepsilon)-f_0(\varepsilon-\mu^{\text{eff}}_{\sigma},T^{\text{eff}}_{\sigma})]=0,
\end{eqnarray}
where we have assumed that the DOS of the N reservoir is equal to that of the N layer.
We wish to warn the reader that the words ``temperature'' and ``chemical potential'' have to be taken in a loose sense, especially when the distributions are very different from equilibrium functions.
They are merely two parameters which grasp important characteristic properties of the distributions, related, namely, to particle and heat transport.

A general expression for the effective temperature can be easily derived through the Sommerfeld expansion \cite{pekola}, obtaining
\begin{equation}
T^{\text{eff}}_{\sigma}=\frac{\sqrt{6}}{\pi k_{\text{B}}} 
\sqrt{ \int_{-\infty}^{\infty}  d\varepsilon \; \varepsilon [f_{\sigma}(\varepsilon)-f_0(\varepsilon-\mu^{\text{eff}}_{\sigma},T=0)
]}
\label{efftemp}
\end{equation}
where
\begin{equation}
\mu^{\text{eff}}_{\sigma}=\int_{-\infty}^{\infty}  d\varepsilon
[f_{\sigma}(\varepsilon)-\theta(\varepsilon)],
\end{equation}
and $\theta(\varepsilon)$ is the Heaviside step function. Equation (\ref{efftemp}) yields the true spin-dependent electron temperature in (quasi)equilibrium. Furthermore, in the present FS-I-N-I-SF system, $T^{\text{eff}}_{\sigma}$ turns out to depend on the strength of electron-electron interaction (i.e., on $\mathcal{K}_{\text{coll}}$) as we shall show in the following, due to heat exchange with FS reservoirs with nonconstant DOS \cite{RMP,pekola}.
\begin{figure}[t!]
\includegraphics[width=\columnwidth,clip]{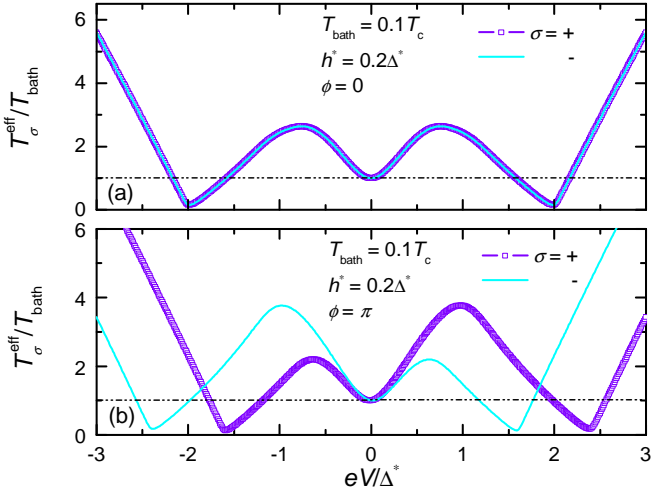}
\caption{ (color online) (a) Nonequilibrium spin-dependent electron effective temperature $T_{\sigma}^{\text{eff}}$ vs $V$ calculated for $\phi=0$ at $T_{\text{bath}}=0.1T_{\text{c}}$ and $h^*=0.2\Delta^*$. (b) $T_{\sigma}^{\text{eff}}$ vs $V$ calculated for $\phi=\pi$ at $T_{\text{bath}}=0.1T_{\text{c}}$ and $h^*=0.2\Delta^*$.}
\label{teff}
\end{figure}

In Figs.~\ref{teff}(a) and (b) we plot the calculated effective temperatures in the full nonequilibrium limit (i.e., for $\mathcal{K}_{\text{coll}}=0$), normalized with respect to the bath temperature ($T_{\text{bath}}=0.1T_{\text{c}}$), as a function of the applied bias voltage $V$ for $\phi=0$ and $\phi=\pi$, respectively.
In the former case, there is no spin dependence and $T^{\text{eff}}_{\pm}$ is an even function of the bias voltage.
Starting from the equilibrium condition (i.e., at $V=0$ where the $T^{\text{eff}}_{\pm}=T_{\text{bath}}$), the temperature first increases reaching a maximum around $eV/\Delta^*=0.8$ and thereafter decreases down to the minimum around $eV/\Delta^*=2$.
The initial increase is the ``anomalous heating'' due to the presence of a finite DOS within the superconducting gap \cite{RMP,pekola}, while the minimum reflects the usual electron ``cooling'' which is maximized for voltages around twice the value of the gap \cite{RMP} (see the discussion in Sec. \ref{non}).
Such features are present also in the absence of an exchange field, with negligible quantitative differences.
By contrast, the effect of the exchange field is manifest in the antiparallel configuration.
Indeed, the effective temperatures are different for the two spin species, though related according to the relation $T^{\text{eff}}_{\sigma}(-V)=T^{\text{eff}}_{-\sigma}(V)$.
In particular, minima are shifted by an amount equal to $2h^*$ towards higher (lower) voltages for spin up (down) electrons.
This fact can also be understood through the schemes shown in Fig.~\ref{schemeDOS}, as due to the shift, introduced by the exchange field, of the DOS of the two superconductors in opposite directions.
For spin up electrons, maximum cooling occurs for $eV=2(\Delta^*-h^*)$ [see Fig.~\ref{schemeDOS}(b)] and for $eV=-2(\Delta^*+h^*)$ [see Fig.~\ref{schemeDOS}(d)]. For spin down electrons maximum cooling occurs at opposite voltages.
The position and amplitude of maxima turns out to be function of the exchange field, as well as of the smearing parameter $\Gamma$. Notably, as shown in Fig. \ref{teff}(b), the spin-dependent effective temperatures can be largely different upon voltage biasing the  structure.

\begin{figure}[t!]
\includegraphics[width=\columnwidth,clip]{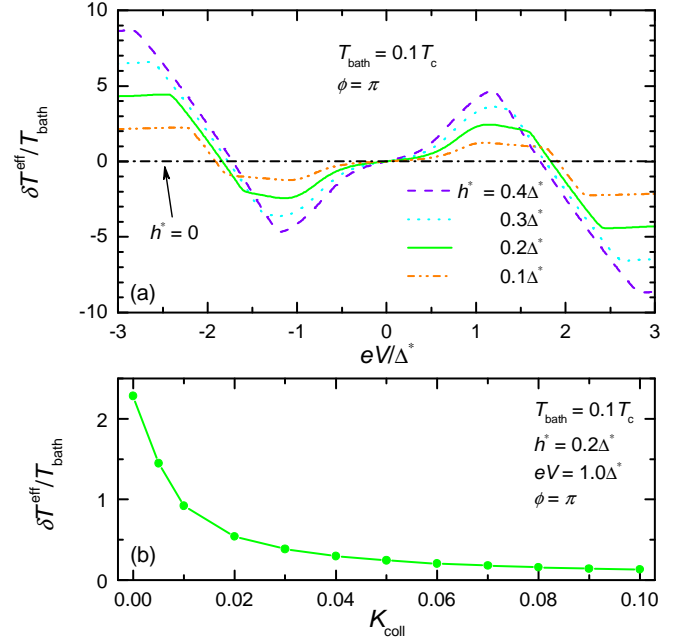}
\caption{ (color online) (a) Difference of the nonequilibrium spin-dependent electron effective temperatures $\delta T^{\text{eff}}$ vs $V$ calculated for different $h^*$ values at $T_{\text{bath}}=0.1T_{\text{c}}$ and $\phi=\pi$. (b) $\delta T^{\text{eff}}$ vs $\mathcal{K}_{\text{coll}}$ calculated at $eV=1.0\Delta^*$ for $T_{\text{bath}}=0.1T_{\text{c}}$, $h^*=0.2\Delta^*$, and $\phi=\pi$.}
\label{teffrelax}
\end{figure}

In Fig.~\ref{teffrelax}(a) we plot the nonequilibrium effective temperature difference $\delta T^{\text{eff}}=T^{\text{eff}}_+-T^{\text{eff}}_-$ normalized to the bath temperature ($T_{\text{bath}}=0.1T_{\text{c}}$) versus $V$ at $\phi=\pi$ and for different values of the exchange field $h^*$.
$\delta T^{\text{eff}}$ is odd in the applied voltage and is more pronounced for larger values of $h^*$.
For positive values of $V$, the maximum (as large as 500 \% at this bath temperature) corresponds to the anomalous heating (occurring around $eV/\Delta^*\simeq 1$), while the minimum occurs for $eV/\Delta^*\gtrsim 2$, and moves to higher values as $h^*$ increases.

The effect of electron-electron inelastic collisions is addressed in Fig.~\ref{teffrelax}(b)  which shows
$\delta T^{\text{eff}}/T_{\text{bath}}$ at $\phi=\pi$ as a function of the collision strength $\mathcal{K}_{\text{coll}}$ for $eV=1.0\Delta^*$, $h^*=0.2\Delta^*$, and $T_{\text{bath}}=0.1T_{\text{c}}$.
We find that a dramatic effect of electron-electron interaction, that leads to a strong suppression of the effective temperature difference on the scale of $\mathcal{K}_{\text{coll}}\simeq 0.1$. With the material parameters given in Sec. \ref{sec:coll} this would correspond to a N region with $t_{\text{N}}\approx 470$ nm.
\begin{figure}[t!]
\includegraphics[width=\columnwidth,clip]{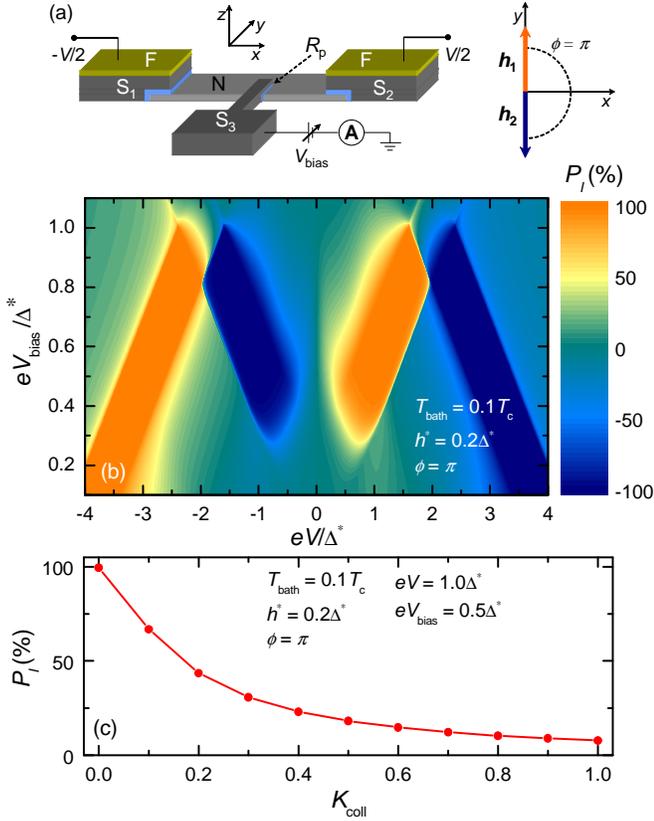}
\caption{ (color) (a) Sketch of a possible setup for the implementation of a spin-polarized current source. An additional superconducting electrode (S$_3$) is coupled to the N region through a tunnel junction of resistance $R_{\text{p}}$. The exchange fields in S$_{1,2}$ are arranged in the antiparallel configuration. Spin-polarized current can be extracted by biasing the S$_3$ terminal with $V_{\text{bias}}$. 
(b) Contour plot of the nonequilibrium current polarization $P_I$ vs $V$ and $V_{\text{bias}}$ at $T_{\text{bath}}=0.1T_{\text{c}}$ and $h^*=0.2\Delta^*$.
(c) $P_I$ vs $\mathcal{K}_{\text{coll}}$ calculated $eV=1.0\Delta^*$ and $eV_{\text{bias}}=0.5\Delta^*$ for $T_{\text{bath}}=0.1T_{\text{c}}$, $h^*=0.2\Delta^*$, and $\phi=\pi$.}
\label{Prelax}
\end{figure}

The possibility to have access to different spin-dependent electronic temperatures suggests that we investigate the potential for the implementation of spintronic devices. 
In particular, we conceive a nanostructure like that shown in Fig. \ref{Prelax}(a) where an additional voltage-biased \emph{superconducting} electrode (S$_3$) is tunnel coupled to the N region through a junction of resistance $R_{\text{p}}\gg R_{\text{t}}$ \cite{rp}, while the exchange fields in S$_{1,2}$ are arranged in the antiparallel configuration ($\phi=\pi$). 
The presence of a superconducting extraction lead is crucial, since in the tunneling process the quasiparticle current through S$_3$ will depend exponentially on the electron temperature in N \cite{RMP}. 
On one hand, the setup considered allows direct measurement of the spin-dependent electron temperatures and S$_3$ may act as a thermometer \cite{RMP}.
S$_3$ provides, in fact, access to the whole distribution functions $f_{\sigma}(\varepsilon)$ from the voltage-dependent differential conductance of the NIS$_3$ junction \cite{pothier}. 
On the other hand, upon biasing the S$_3$ electrode with $V_{\text{bias}}$, the existence of different spin-dependent temperatures in the normal metal region yields a finite current polarization $P_I$ defined in the usual way as
\begin{equation}
P_I(V,V_{\text{bias}},h^*)=\frac{I_+^{\text{S}_3}(V,V_{\text{bias}},h^*) -I_-^{\text{S}_3}(V,V_{\text{bias}},h^*)}{I_+^{\text{S}_3}(V,V_{\text{bias}},h^*) +I_-^{\text{S}_3}(V,V_{\text{bias}},h^*)},
\end{equation}
where
\begin{equation}
\begin{split}
I_{\sigma}^{\text{S}_3}(V,V_{\text{bias}},h^*)=\frac{1}{eR_{\text{p}}}\int d\varepsilon\mathcal{N}^{\text{S}_3}(\varepsilon)\,\,\,\,\,\,\,\,\,\,\,\,\,\,\,\,\,\,\,\,\,\,\,\,\,\,\,\,\,\,\,\,\,\,\,\,\,\, \\
\times[f_{\sigma}(\varepsilon,V,h^*)-f_0(\varepsilon+eV_{\text{bias}})],
\end{split}
\end{equation}
and $\mathcal{N}^{\text{S}_3}(\varepsilon)$ is the normalized DOS of S$_3$. In the following we assume for simplicity that $\mathcal{N}^{\text{S}_3}(\varepsilon)$ is identical to the density of states of $\text{S}_{\text{1,2}}$ in the absence of an exchange field (i.e., $h^*$=0).

Figure \ref{Prelax}(b) shows the calculated nonequilibrium $P_I$ as a function of $V$ and $V_{\text{bias}}$, for $T_{\text{bath}}=0.1T_{\text{c}}$ and $h^*=0.2\Delta^*$. For easily attainable values of $V$ and $V_{\text{bias}}$ pure (i.e., $100\%$) spin-polarized current of both the parallel and antiparallel spin species can be achieved. Furthermore, we note that $|P_I|$ largely exceeds $50\%$ over a wide region in the ($V,\,V_{\text{bias}}$) plane. 
It is worthwhile to note that replacing the S$_3$ terminal with a N metal reservoir would completely hinder the extraction of a finite spin-polarized current for $\phi=\pi$. This stems from the insensitivity of the tunneling current in a NIN junction  to the N region temperature. 
The $\phi=0$ case for a N lead connected to a similar setup was analyzed in Ref. \onlinecite{giazotto05}. 

The role of electron-electron inelastic relaxation is displayed in Fig.~\ref{Prelax}(c) which shows $P_I$ versus $\mathcal{K}_{\text{coll}}$ at $eV=\Delta^*$ and $eV_{\text{bias}}=0.5\Delta^*$, for $T_{\text{bath}}=0.1T_{\text{c}}$ and $h^*=0.2\Delta^*$.
We find that the suppression of the current polarization occurs on a larger range of values of $\mathcal{K}_{\text{coll}}$ with respect to electron effective temperatures difference [note the different scale for the horizontal axis of Fig.~\ref{teffrelax}(b) and \ref{Prelax}(c)].
At $\mathcal{K}_{\text{coll}}=0.1$, for example, $P_I$ is reduced only by about $35\%$.
This behavior is not surprising: it originates, as mentioned above, from the fact that the current extracted from S$_3$ is exponentially sensitive to $T^{\text{eff}}_{\sigma}$.
As a consequence, even a small temperature difference between spin species yields a large current difference, and gives rise to a sizable $P_I$ vs $\mathcal{K}_{\text{coll}}$ characteristic.

We shall further comment the impact of relaxation in this setup as compared to that of Fig. 1 by inspecting Fig. 14(c) and Figs. 11(b-d). In particular they show that while electron-electron interaction is crucial in suppressing $P_I$ in the former case, it is much less important in the setup of Fig. 1. The reason stems from the fact that while in the present case the electric current is spin-dependent only thanks to presence of spin-dependent distribution functions in the N region (as expressed by Eq. 24), in the setup of Fig. 1 the spin selectivity originates from both the distribution functions $f_{\sigma}$  and the spin-dependent superconducting DOS (see Eqs. 10 and 11). From this follows that, in the first setup, current polarization will completely vanish at quasiequilibrium (where the quasiparticle distribution functions result to be identical and spin-independent), while in the second system spin polarization will persists also for identical thermal distribution functions owing to the additional spin selectivity provided by spin-split DOS in the superconductors.

\section{Conclusions}
\label{sec:conc}
In this paper we have analyzed the nonequilibrium spin-dependent transport properties in superconductor-normal metal tunnel nanostructures, where the superconductors present a proximity-induced {\em effective} exchange field (equivalent to that of a superconductor in a magnetic field).
The latter is due to the proximity of a ferromagnetic thin layer which exerts, under appropriate conditions, a nonlocal influence.
We have computed the quasiparticle distributions of a normal metal layer tunnel coupled to two superconductors with non-collinear exchange fields.
In the full nonequilibrium limit, i.e., in the absence of any inelastic relaxation mechanisms, the distribution functions depend on spin and display unusual features, such as population inversion and double-step shape, depending on the bias voltage applied to the superconductors. 
Spin-dependence persists also in the presence of inelastic Coulomb interaction, which produces a smoothing of the sharp features and tends to drive electrons into the quasiequilibrium regime, where the energy relaxation is strong enough to allow electrons to thermalize.
This interesting behavior is reflected in the current-voltage characteristic, which shows a strong dependence on the relative orientation of the exchange fields existing in the superconductors.
Notably, even in the quasiequilibrium regime, a huge tunnel magnetoresistance ratio and a complete current spin-polarization were found over a wide range of bias voltages and for realistic parameters.
The impact of the exchange field as well as of the presence of subgap states in the DOS of the superconductors have been addressed.

We have characterized the out-of-equilibrium distribution functions  through an ``effective temperature'' and an ``effective chemical potential'', defined in an operative sense.
We have found that such effective temperatures are strongly spin-dependent for anti-parallel exchange fields, the relative temperature difference being as high as $500\%$.
Furthermore, we have discussed the possibility of producing spin-polarized currents by coupling the N region to an additional superconducting lead, finding that $100\%$ spin-polarization is realistically achievable.
This effect is fairly robust against the occurrence of inelastic electron-electron collisions.

We shall finally comment on some possible applications of the structures here presented.
An immediate first application of this system is the implementation of storage cell elements, thanks to the very large  TMR values [see Fig.~\ref{tF}(a)]. Magnetic-field-controlled current switches can be envisioned as well [see Fig.~\ref{tS}(a)].
Importantly, power dissipation is intrinsically limited owing to the small currents driven through  NIS junctions. For example, assuming $R_{\text{t}}=10^3\,\Omega$ and aluminum (Al) electrodes at $T=0.1T_{\text{c}}\approx  0.12$ K, a dissipated power in the range of $10^{-15}\ldots10^{-12}$ W can be achieved for $2(\Delta^*-h^*)/e<|V|<2\Delta^*/e$. This makes this setup attractive for low-dissipation cryogenic applications. 
In light of a realistic implementation, ferromagnetic alloys such as Cu$_{1-x}$Ni$_x$ \cite{ryazanov} or Pd$_{1-x}$Ni$_x$ \cite{kontos} (which allow fine tuning of $h$ through a proper choice of $x$) are promising candidates. 
For example, in Pd$_{1-x}$Ni$_x$ alloy with $x=0.1$, $h\simeq 10$ meV resulting in $\xi_{\text{F}}\approx 5$ nm \cite{kontos}. By choosing Al as S electrodes (with $\Delta\simeq 200\,\mu$eV and $\xi_{\text{S}}\approx 300$ nm \cite{romijn}) it turns out that $h^*$ in the range $\sim0.2\Delta^* ... 0.5\Delta^*$ can be achieved. In such nanostructures the bias voltage con be fed through outer normal metal electrodes, tunnel coupled to the ferromagnetic layers in order to prevent depolarization of the F electrodes. This would result in adding in series extra resistances which could be easily engineered in order to minimize the induced correction to both tunnel magnetoresistance and operating voltage.

\section{Acknowledgments}
Partial financial support from the EU funded HYSWITCH, NanoSciERA ``NanoFridge'' and RTNNANO projects is acknowledged.


\begin{thebibliography}{99}
\bibitem{kopnin}
	{\it Theory of Nonequilibrium Superconductivity}, N. B. Kopnin (Clarendon, Oxford, 2001).
\bibitem{samuelsson00}
	P. Samuelsson, J. Lantz, V. S. Shumeiko, and G. Wendin, Phys. Rev. B \textbf{62}, 1319 (2000).
\bibitem{pothier}
        H. Pothier, S. Gu\'eron, N. O. Birge, D. Esteve, and M. H. Devoret,
        Phys. Rev. Lett. \textbf{79}, 3490 (1997).
\bibitem{vanwees}
        B. J. van Wees, K.-M. H. Lenssen, and C. J. P. M. Harmans, Phys. Rev. B \textbf{44}, 470 (1991).
\bibitem{samuelsson97}
	P. Samuelsson, V. S. Shumeiko, and G. Wendin, Phys. Rev. B \textbf{56}, R5763 (1997).
\bibitem{shumeiko}
	V. S. Shumeiko, G. Wendin, and E. N. Bratus’, Phys. Rev. B \textbf{48}, 13129 (1993).
\bibitem{gorelik}
	L. Y. Gorelik, V. S. Shumeiko, R. I. Shekhter, G. Wendin, and M. Jonson, Phys. Rev. Lett. \textbf{75}, 1162 (1995).
\bibitem{volkov95}
	A. F. Volkov, Phys. Rev. Lett. \textbf{74}, 4730 (1995).
\bibitem{volkov97}
	A. F. Volkov and H. Takayanagi, Phys. Rev. B \textbf{56}, 11184 (1997).
\bibitem{wilhelm}
	F. K. Wilhelm, G. Sch\"on, and A. Zaikin, Phys. Rev. Lett. \textbf{81}, 1682 (1998).
\bibitem{yip}
	S. K. Yip, Phys. Rev. B \textbf{58}, 5803 (1998).
\bibitem{morpurgo98}
	A. F. Morpurgo, T. M. Klapwijk, and B. J. van Wees, Appl. Phys. Lett. \textbf{72}, 966 (1998).
\bibitem{baselmans99} 
	J. J. A. Baselmans, A. F. Morpurgo, B. J. van Wees, and T. M. Klapwijk,  Nature \textbf{397}, 43 (1999).
\bibitem{giazotto03}
	F. Giazotto, F. Taddei, T. T. Heikkil\"a, R. Fazio, and F. Beltram, Appl. Phys. Lett. \textbf{83}, 2877 (2003).
\bibitem{SINIS}
	F. Giazotto, T. T. Heikkil\"a, F. Taddei, R. Fazio, J. P. Pekola, and F. Beltram, Phys. Rev. Lett. \textbf{92}, 137001 (2004).
\bibitem{JT}
	F. Giazotto and J. P. Pekola, J. Appl. Phys. \textbf{97}, 023908 (2005).
\bibitem{laakso}
	M. A. Laakso, P. Virtanen, F. Giazotto, and T. T. Heikkil\"a, Phys. Rev. B \textbf{75}, 094507 (2007).
\bibitem{savin04}
	    A. M. Savin, J. P. Pekola, J. T. Flyktman, A. Anthore, and F. Giazotto, Appl. Phys. Lett. \textbf{84}, 4179 (2004).
\bibitem{SER}
	J. P. Pekola, F. Giazotto, and O.-P. Saira, Phys. Rev. Lett. \textbf{98}, 037201 (2007).
\bibitem{HT}
	O.-P. Saira, M. Meschke, F. Giazotto, A. M. Savin, M. M\"ott\"onen, and J. P. Pekola, Phys. Rev. Lett. \textbf{99}, 027203 (2007).
\bibitem{phystoday}
	J. P. Pekola, R. Schoelkopf, and J. Ullom, Phys. Today \textbf{57}, No. 5, 41 (2004).
\bibitem{RMP}
	F. Giazotto, T. T. Heikkil\"a, A. Luukanen, A. M. Savin, and J. P. Pekola, Rev. Mod. Phys. \textbf{78}, 217 (2006).
\bibitem{taka} S. Takahashi, H. Imamura, and S. Maekawa, Phys. Rev. Lett. \textbf{82}, 3911 (1999).
\bibitem{maekawa} S. Maekawa, S. Takahashi, and H. Imamura, J. Phys. D: Appl. Phys. \textbf{35}, 2452 (2002).
\bibitem{tser} Y. Tserkovnyak and A. Brataas, Phys. Rev. B \textbf{65}, 094517 (2002).
\bibitem{johansson} J. Johansson, V. Korenivski, D. B. Haviland, and A. Brataas,
	Phys. Rev. Lett. \textbf{93}, 216805 (2004).
\bibitem{taka2} S. Takahashi, T. Yamashita, T. Koyama, and S. Maekawa, J. Appl. Phys. \textbf{89}, 7505 (2001).
\bibitem{bobkova}
	I. V. Bobkova and A. M. Bobkov, Phys. Rev. B \textbf{74}, R220504 (2006).
\bibitem{belzig00}
	W. Belzig, A. Brataas, Yu. V. Nazarov, and G. E. W. Bauer, Phys. Rev. B \textbf{62}, 9726 (2000).
\bibitem{giazotto05}
	F. Giazotto, F. Taddei, R. Fazio, and F. Beltram, Phys. Rev. Lett. \textbf{95}, 066804 (2005).
\bibitem{giazotto06-2}
	F. Giazotto, F. Taddei, R. Fazio, and F. Beltram, Appl. Phys. Lett. \textbf{89}, 022505 (2006).
\bibitem{buzdin}
	See A. I. Buzdin, Rev. Mod. Phys. \textbf{77}, 935 (2005); F. S. Bergeret, K. B. Efetov, and A. Volkov, Rev. Mod. Phys. \textbf{77}, 1321 (2005), and references therein.
\bibitem{bergeret}
	F. S. Bergeret, A. F. Volkov, and K. B. Efetov, Phys. Rev. Lett. \textbf{86}, 3140 (2001).
\bibitem{spinorbit}
	P. M. Tedrow and R. Meservey, Phys. Rev. Lett. \textbf{27}, 919 (1971).
\bibitem{meservey}
	R. Meservey and P. M. Tedrow, Phys. Rep. \textbf{238}, 173 (1994).

\bibitem{pekola}
	J. P. Pekola, T. T. Heikkil\"a, A. M. Savin, J. T. Flyktman, F. Giazotto, and F. W. J. Hekking, Phys. Rev. Lett. \textbf{92}, 056804 (2004).
\bibitem{Dynes}
	R. C. Dynes, J. P. Garno, G. B. Hertel, and T. P. Orlando, Phys. Rev. Lett. \textbf{53}, 2437 (1984).
\bibitem{heslinga}
	D. R. Heslinga and T. M. Klapwijk,
	Phys. Rev. B \textbf{47}, 5157 (1993).
\bibitem{alt}
        B. L. Altshuler and A. G. Aronov, in \emph{Electron-Electron 
        Interactions in Disordered Systems}, edited by A. L. Efros  and M. Pollak 
        (Elsevier, Amsterdam, 1985).
\bibitem{kaminski}
        A. Kaminski and L. I. Glazman, Phys. Rev. Lett. 
        \textbf{86}, 2400 (2001).
\bibitem{anthore}
        A. Anthore, F. Pierre, H. Pothier, and D. Esteve,
        Phys. Rev. Lett. \textbf{90}, 076806 (2003).
\bibitem{nagaev}
        K. E. Nagaev, Phys. Rev. B \textbf{52}, 4740 (1995).
\bibitem{KL}
	M. Yu. Kuprianov and V. F. Lukichev, Zh. Eksp. Teor. Fiz. \textbf{94}, 139 (1988) [Sov. Phys. JETP \textbf{67}, 1163 (1988)].      
\bibitem{alt2}
        B. L. Altshuler and A. G. Aronov, Zh. Eksp. Teor. Fiz.
        \textbf{75}, 1610 (1978) [Sov. Phys. JETP \textbf{48}, 812 (1978)]. 
\bibitem{kamenev}
        A. Kamenev and A. Andreev, Phys. Rev. B 
        \textbf{60}, 2218 (1999).
\bibitem{huard}
	B. Huard, A. Anthore, F. Pierre, H. Pothier, N. O. Birge, and D. Esteve, Solid State Commun. \textbf{131}, 599 (2004).

\bibitem{kupriyanov}
	M. Yu. Kupriyanov, A. Brinkman, A. A. Golubov, M. Siegel, and H. Rogalla, Physica C \textbf{326-327}, 16 (1999).

\bibitem{tinkham}
	M. Tinkham, \emph{Introduction to Superconductivity} (Dover Publications, New York, 1996).


\bibitem{jedema}
	F. J. Jedema, A. T. Filip, and B. J. van Wees, Nature (London) \textbf{410}, 345 (2001).

\bibitem{johnson}
	M. Johnson, Phys. Rev. Lett. \textbf{70}, 2142 (1993).


\bibitem{tesiheikkila} T. T. Heikkil\"a, \emph{Superconducting proximity effect in mesoscopic metals}, PhD thesis, HUT Helsinki 2002, Finland.

\bibitem{rp}
The constraint $R_{\text{p}}\gg R_{\text{t}}$ ensures the nonequilibrium condition in the N region not to be appreciably perturbed by the current extraction from the superconducting terminal S$_3$.

\bibitem{ryazanov}
	V. V. Ryazanov, V. A. Oboznov, A. Yu. Rusanov, A. V. Veretennikov, A. A. Golubov, and J. Aarts, Phys. Rev. Lett. \textbf{86}, 2427 (2001).

\bibitem{kontos}
	T. Kontos, M. Aprili, J. Lesueur, and X. Grison, Phys. Rev. Lett. \textbf{86}, 304 (2001).

\bibitem{romijn}
	J. Romijn, T. M. Klapwijk, M. J. Renne, and J. E. Mooij, Phys. Rev. B. \textbf{26}, 3648 (1982).


\end{thebibliography}
\end{document}